\documentstyle[epsf,rotate]{aa}

\newcommand{\kms}{\mbox{\,km\,s$^{-1}$}}
\newcommand{\Msun}{\mbox{M$_{\odot}$}}
\newcommand{\Rsun}{\mbox{R$_{\odot}$}}
\newcommand{\Lsun}{\mbox{L$_{\odot}$}}

\newcommand{\ltsimeq}{\raisebox{-0.6ex}{$\,\stackrel 
        {\raisebox{-.2ex}{$\textstyle <$}}{\sim}\,$}} 
\newcommand{\gtsimeq}{\raisebox{-0.6ex}{$\,\stackrel
        {\raisebox{-.2ex}{$\textstyle >$}}{\sim}\,$}}

\begin{document}

\title{The rotation speed of the companion star in 
V395 Car (=2S0921--630)  }

\author{T.~Shahbaz$^{1}$ 
\thanks{Based on observations made at the European Southern Observatory,
La Silla, Chile}
\and E.~Kuulkers$^{1}$ \and P.A.~Charles$^{1}$
\and F.~van der Hooft$^{2}$ \and J.~Casares$^{3}$ 
\and J.~van Paradijs$^{2}$}

\offprints{T.~Shahbaz (tsh@astro.ox.ac.uk)}

\institute{$^{1}$Department of Astrophysics, Oxford University, Keble Road,
Oxford, OX1 3RH, UK \\
$^{2}$Astronomical Institute ``Anton Pannekoek'', University of Amsterdam
and Center for High Energy Astrophysics, Kruislaan, 403, 1098 SJ Amsterdam, 
The Netherlands \\
$^{3}$Instituto de Astrof\'\i{}sica de Canarias 38200 La Laguna, 
Tenerife, Spain}

\thesaurus{2(08.02.3; 08.06.3; 08.09.2: V395 Car; 08.14.1)}

\date{Received ?? ?? 1997 / Accepted ?? ?? 1997}

\maketitle

\markboth{T.~Shahbaz et al: V395 Car}{}


\begin{abstract}
We have obtained high resolution optical spectroscopy of the accretion
disc corona source 2S0921--630 (=V395 Car) that constrains the spectral
type of the companion star to be most probably K0$\sc iii$. At 6500\AA\
the secondary star is found to contribute $\sim$ 25 \% to the observed
flux. The absorption line spectrum of the K0$\sc iii$ companion star is
broadened (compared to template stars) by a projected rotational velocity
of 65$\pm$9\kms (1-$\sigma$).

\end{abstract}


\section{Introduction}

In spite of its low luminosity, 2S0921-630 has proved to be an
interesting and important X-ray source.  Discovered by SAS-3 and
identified with a $\sim$15$^m$ blue star, V395 Car (Li et al, 1978),
subsequent studies (e.g. Branduardi-Raymont et al, 1983) showed that the
spectrum was dominated by strong HeII $\lambda$4686 and Balmer emission,
a characteristic of low-mass X-ray binaries (LMXBs).  However, 2S0921-630
has an unusually low $L_X/L_{opt}$ of $\sim$1 compared to most LMXBs,
which was explained by the presence of partial optical and X-ray eclipses
(Mason et al, 1987). This demonstrated that 2S0921-630 was a high
inclination, {\it accretion-disc corona} (ADC) source.  Only a handful of
these are known (e.g. White et al., 1995) in which the compact object
is permanently obscured from our view by the accretion disc, and hence
the observed X-rays are scattered into our line-of-sight by a hot corona
of gas above and below the disc.  By implication, the intrinsic X-ray
luminosity is much higher.

However, what is remarkable about 2S0921-630 is its long period.  The
other ADC sources are 2A1822-371, \, 4U2129+47 and 4U2127+19, which have
relatively short orbital periods (5.6, 5.2 and 17.1 hrs respectively).
Optical photometry and spectroscopy (Cowley et al., 1982 and
Branduardi-Raymont et al., 1983) have indicated that V395 Car has a much
longer orbital period of 9.02 days, with kinematic properties implying
that it is located in the halo at a distance of $\sim$10 kpc (Cowley et
al., 1982). The secondary star must then be evolved and intrinsically
luminous in its own right, making the system very similar to the
well-known halo giant Cyg X-2 (Casares et al., 1997; Orosz \& Kuulkers,
1998).  With its high inclination V395 Car is thus one of those rare
LMXBs (along with Cyg X-2 and Her X-1) in which the secondary should be
visible in spite of the presence of a luminous disc, and hence high
resolution optical spectroscopy should reveal both the spectral type of
the secondary and its rotation speed, thereby providing significant
constraints on the system masses.  Here we report the first results from
such a study.

\begin{table}
\centering
\small{
\caption{Log of observations}
\begin{tabular}{lcc}\hline
\hline\noalign{\smallskip}
Date        &    UT    &  Exp. time \\ \hline
            &          &            \\
28/05/1998  & 23:41:48 & 1200s      \\
29/05/1998  & 00:10:52 & 1200s      \\
29/05/1998  & 00:32:55 & 1200s      \\ 
            &          &            \\ \hline
\noalign{\smallskip}
\hline
\end{tabular}
}
\end{table}

\section{Observations and data reduction}

High resolution optical spectra of V395 Car were obtained on 1998 May
28/29 with the 3.5-m New Technology Telescope (NTT) at the European
Southern Observatory (ESO) in Chile using the ESO Multi Mode Instrument
(EMMI). We used the red arm with an order-separating OG 530 filter and
grating \#6 which gave a dispersion of 0.31~\AA\ per pixel. The TEK
2048$\times$2048 CCD was used, binned by a factor two in the spatial
direction in order to reduce the readout noise. The dispersion direction
was not binned. Very good seeing allowed us to use a slit width of
$0\farcs8$ which resulted in a spectral resolution of 0.83~\AA. We took
3$\times$1200sec exposures of V395 Car (see Table 1) and Cu-Ar arc
spectra were taken for wavelength calibration. In addition we observed
template field stars with a variety of spectral types, whose rotational
velocities are much less than the resolution of our data.

The data reduction and analysis was performed using the Starlink {\sc
figaro} package, the {\sc pamela} routines of K.\,Horne and the {\sc
molly} package of T.\,R.\ Marsh. Removal of the individual bias signal
was achieved through subtraction of the mean overscan level on each
frame. Small scale pixel-to-pixel sensitivity variations were removed
with a flat-field frame prepared from observations of a tungsten lamp.
One-dimensional spectra were extracted using the optimal-extraction
algorithm of Horne (1986), and calibration of the wavelength scale was
achieved using 5th order polynomial fits which gave an rms scatter of
0.02~\AA. The stability of the final calibration was verified with the OH
sky line at 6562.8\AA\ whose position was stable to within 0.1 \AA.

\begin{picture}(0,0)(0,0)
  \put(0,0){\includegraphics{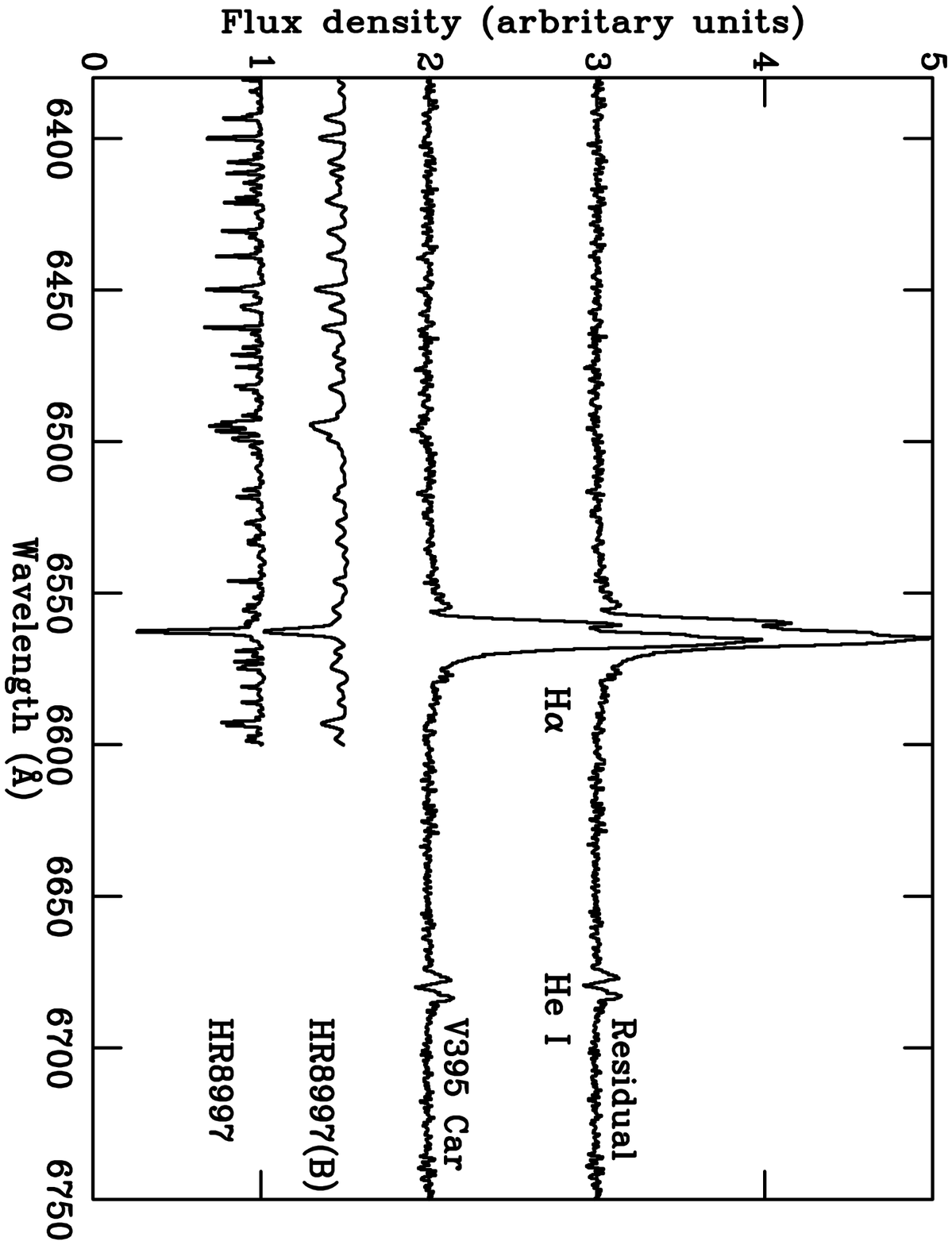}}
\end{picture}

\begin{figure}
  \vspace{8.0cm}
  \caption{ The results of the optimal subtraction. From bottom to top:
  the template K0$\sc iii$ star (HR8997), the template star broadened by
  64\kms, the variance-weighted average spectrum of V395 Car and the
  residual spectrum of V395 Car after subtracting the template star times
  $f$=0.25.The spectra have been normalized and shifted vertically for
  clarity.}
\end{figure}

\section{The V395 Car spectrum}

In Fig. 1 we show the variance-weighted average of our V395 Car spectra,
which has a signal-to-noise ratio of about 40 in the continuum. The most
noticeable features are the double peaked H$\alpha$ and He$\sc i$
6678\AA\ emission lines. The double peaked nature of the emission lines
is characteristic of a high inclination X-ray binary (Horne \& Marsh 1986).
This is consistent with the observed partial eclipse in the optical and
X-rays (Mason et al., 1987).

\section{The spectral type and rotational broadening of the companion star}

We determine the spectral type of the companion star by minimizing the
residuals after subtracting different template star spectra from the
Doppler-corrected average spectrum. This method is sensitive to the
rotational broadening $v\sin\,i$ and the fractional contribution of the
companion star to the total flux ($f$; 1-$f$ is the ``veiling factor'').
The template stars we use are in the spectral range F2--K2 $\sc iii$ and
were obtained during this observing run but also from previous runs at La
Palma and with comparable dispersion.

First we determined the velocity shift of the individual spectra of V395
Car with respect to each template star spectrum by the method of
cross-correlation (Tonry \& Davis 1979). The V395 Car spectra were then
interpolated onto a logarithmic wavelength scale (pixel size 14.5 \kms)
using a $\sin\,x/x$ interpolation scheme to minimize data smoothing
(Stover et al. 1980). The spectra of V395 Car were then Doppler-averaged
to the rest frame of the template star.

In order to determine the rotational broadening $v\sin\,i$ we follow the
standard procedure described by Marsh et al., (1994). Basically we
subtracted a constant representing the fraction of light from the
template star, multiplied by a rotationally broadened version of that
template star. We broadened the template star spectrum from 0 to 100
km~s$^{-1}$ in steps of 1 km~s$^{-1}$ using the Gray rotation profile
(Gray 1992). We then performed an optimal subtraction (Marsh et al.,
1994) between the broadened template and averaged V395 Car spectra. The
optimal subtraction routine adjusts the constant to minimize the residual
scatter between the spectra. The scatter is measured by carrying out the
subtraction and then computing the $\chi^{2}$ between this and a smoothed
version of itself. The constant, $f$, represents the fraction of light
arising from the template spectrum, i.e. the secondary star. The optimal
values of $v\sin\,i$ and $f$ are obtained by minimising $\chi^{2}$. 
The above analysis was performed in the spectral ranges
6380--6520 \AA\ which excludes H$\alpha$ and He$\sc i$ 6678\AA . This was
the only region common to all the templates stars and V395 Car. A linear
limb-darkening coefficient of 0.60 was used (Al-Naimiy 1978).

Using the template stars covering a range in spectral type (F2--K2$\sc
iii$) we found $v\sin\,i$ to be in the range 58--71 \kms. The minimum
$\chi^{2}$ occured at spectral type K0 with a $v\sin\,i$ of 64$\pm$ 9
\kms (1-$\sigma$) and the companion star contributing about 25\% to the
observed flux at $\sim$ 6500 \AA. Fig. 1 shows the results of the optimal
subtraction.
This analysis assumes that the limb-darkening coefficient
appropriate for the radiation in the line is the same as for the
continuum. However, in reality this is not the case; the absorption lines
in early-type stars will have core limb-darkening coefficients much less
than that appropriate for the continuum (Collins \& Truax 1995). In order
to determine the extreme limits for $v\sin~i$ we also repeated the above
analysis for the K0 template star using zero and full limb-darkening. We
found that $v\sin\,i$ changes by 4 \kms.

In order to estimate the systematic effects in estimating the spectral
type and rotational broadening in V395 Car, we performed the same
analysis but now using a template star of known spectral type. We
broadened the target spectrum by 60 \kms, added noise and veiled it by 70
\% to produce a spectrum of comparable quality to that of V395 Car. 
We then repeated the broadening and optimal subtraction procedure using
the same templates stars as was used for the V395 Car analysis, thereby
determining the best fit. We found that with our analysis we were able to 
retrieved the spectral type to within two subclasses 
and the enforced rotational broadening and veiling accurate to 7\%.

\section{Discussion}

\subsection{Nature of the secondary star}

In the ADC sources the observed X-rays are scattered into our
line-of-sight by a hot corona of gas above and below the disc. The
intrinsic X-ray emission is permanently obscured from our view by the
accretion disc and its extended rim. The observation of partial X-ray
eclipses constrains the binary inclination $i$ to lie in the range
75$^\circ$--90$^\circ$ (Mason et al., 1987). Thus, given our observed
projected rotational broadening of the secondary star (64$\pm$9
\kms) and these limits to $i$ we determine the rotation of the
secondary star to lie in the range 55.0--75.6 \kms (1-$\sigma$ limits).

Furthermore, as a ``steady'' X-ray source, the secondary star must fill
its Roche-lobe, and so its rotational velocity is given by

\begin{equation}
v_{\rm rot} = 611 \left[ \frac{M_{1}(1+q)}{P_{hr}}\right] ^{1/3} 
\left( \frac{R_{\rm L2}}{a}\right)~~~~{\rm km~s^{-1}}
\end{equation}

\noindent
where $P_{hr}$ is the orbital period in hrs, $q$ is the binary mass ratio
(=$M_{2}/M_{1}$), $M_{1}$ is the mass of the neutron star and $R_{\rm
L2}/a$ is the Roche-lobe radius of the secondary and depends only on $q$
(Eggleton 1983). For a given mass for the compact object, $M_{1}$, we can
solve equation (1) for $q$ and hence $M_{2}$, and then also determine
$R_{\rm L2}$. Assuming that the compact object has the mass of a
canonical neutron star, 1.4 \Msun, we find $q$, $M_{2}$ and $R_{\rm L2}$
to lie in the ranges 1.0--2.2, 1.4--3.1 \Msun and 9.7--13.4 \Rsun
respectively.

In a long period X-ray binary, loss of angular momentum via nuclear
expansion of the secondary star drives the mass transfer, provided
$q$\ltsimeq 1.2. In this case, the mass and radius of the secondary star
are constrained to lie in the range 1.4--1.7 \Msun and 9.7--10.3 \Rsun
respectively. Recent theoretical models of King et al. (1997) have
concluded that long period persistent X-ray binaries that contain neutron
stars must have companion masses $M_{2}$\gtsimeq 0.75 \Msun. Our limits
for $M_{2}$ are consistent with this idea. The mean density ($\rho$) of
the secondary star is fixed by the orbital period (Frank et al., 1992). For 2S0921--630 we find $\rho=2.4\times 10^{-3}$~g~cm$^{-3}$,
which implies a K0$\sc iii$ spectral type for the secondary star (Gray
1992). Note that this is consistent with our observed estimate (see Sect.
4).

\subsection{P-Cygni profiles?}

There is evidence for an outflow in 2S0921--630 arising from an accretion
disc wind. The blue spectra of Branduardi-Raymont et al., (1983) show the
Balmer emission lines to have a P-Cygni type profiles, where the blue
wing of the line profile is absorbed. 

As the binary inclination of 2S0921--630 is high, one would expect the
P-Cygni profiles to be stronger at phase 0.5 that at phase 0.0, simply
because at phase 0.5 one sees more of the accretion disc (Note we have
used the usual phase convention i.e. phase 0.0 is defined when the
secondary star is in front of the compact object.) The Balmer lines in
the spectra of Branduardi-Raymont et al., (1983) taken at phase 0.52
(i.e. phase 0.77 in their convention) do indeed show P-Cygni profiles,
Using the orbital ephemeris determined by Mason et al. (1987), we find
that the orbital phase of our NTT spectrum to be 0.03$\pm$0.03. Our
spectrum does not show any evidence for a P-Cygni type profile, which is
what one would expect, in a high inclination binary system.

The spectra of Branduardi-Raymont et al., (1983) taken near orbital phase
0.52 show strong He$\sc i$ 4471\AA\ in absorption. This is clear evidence
for irradiation of the secondary star (c.f. 2A1822-371; Harlaftis et al., 
1997) as late-type stars do not show He$\sc i$ lines. This
suggests that the inner face of the secondary star, facing the compact
object has a mean temperature of $\sim$ 20,000 K i.e. a spectrum of an
early B type star. Note that since our NTT spectrum was taken near phase
0.0 the effects of irradiation will be the least and will most resemble
the ``true'' spectral type of the secondary (compared to spectra taken at
other orbital phases, where the accretion disc light and the effects of
irradiation will contribute more).

\subsection{Comparison with other systems}

We can compare 2S0921--630 with the ADC source 4U2127+119. Both systems
show Balmer P-Cygni type profiles and He$\sc i$ 4471\AA\ in absorption.
The probable high binary mass ratio in 4U2127+119 leads to unstable mass
transfer from the secondary star, resulting in a common envelope (Bailyn
\& Grindlay 1987). In 4U2127+119 the He$\sc i$ absorption line is blue
shifted with respect to the mean velocity of the systemm believed to
arise in a stream of gas leaving the outer Lagrangian point (Bailyn et
al., 1989). If the mass transfer in 2S0921--630 is
unstable, then we would also expect the He$\sc i$ lines to be
blue-shifted and have low non-sinusoidal velocity variations.

It is interesting to note the similarities between 2S0921--630 and Cyg
X--2. Both systems are long period binaries in the halo of the Galaxy,
they are at high inclination angles with evolved secondaries and both
contain neutron stars. [Cyg X-2 must contain a neutron star because of
the observed type$\sc i$ X-ray bursts. For 2S0921--630 we cannot
unequivocally state that it contains a neutron as no bursts have been
seen. However, this may be a result of the natural consenquence of ADC
sources. It should also be noted that our upper limit to $M_{1}$ suggests
that it is a neutron star.] Also 2S0921--630 has a binary mass ratio
$q\sim 1$, which is a factor of 3 higher than that of Cyg X--2 ($q$=0.34;
Casares et al., 1997), suggesting that if the neutron stars in both
systems have similar masses, then the secondary star in 2S0921--630 is
more massive. However, it should be noted that the secondary star in
2S0921--630 is much cooler (K0$\sc iii$; $L_{2}\sim$50 \Lsun) and hence less
luminous than the secondary star in Cyg X--2 (A9$\sc iii$; $L_{2}\sim$200
\Lsun), contrary to what we might have expected given their inferred
masses.  This discrepency can be reconciled if we postulate that we are
seeing 2S0921--630 and Cyg X--2 at very different phases in their
evolution. In Cyg X--2 we are probably seeing the secondary star which is
high up on the giant branch and has lost its outer evelvope due to mass
transfer and/or irradiation, leaving just the hot inner core of what had
been initially a more massive star. Whereas, in 2S0921--630 the secondary
star is not as evolved and so is near the base of the giant branch. Only
detailed stellar evolution calculations will resolve this.

\section{Conclusion}

Using high resolution optical spectra, we estimate the spectral
type of the companion star in 2S0921--630 (=V395 Car) to be a K0$\sc
iii$ star. By optimally subtracting different broadened versions of the
companion star spectrum from the average V395 Car spectrum we determine
the rotational broadening of the companion star to be 64$\pm$9 \kms
(1-$\sigma$) contributing $\sim$ 25\% to the observed flux at
6500\AA. 

\section*{Acknowledgements}

J.C. acknowlegdes support by the Spanish Ministerio de Educacion y
Cultura through the grant FPI-070-97.

\end{document}